\documentclass[fleqn,twoside]{article}
\usepackage{espcrc2}

\usepackage{graphicx}
\usepackage[figuresright]{rotating}

\newcommand{\AmS}{{\protect\the\textfont2
  A\kern-.1667em\lower.5ex\hbox{M}\kern-.125emS}}

\title{Inelastic Diffraction at Heavy Ion Colliders}

\author{Sebastian N. White\address[BNL]{Brookhaven National Laboratory, \\ 
        Upton, N.Y. 11973, USA}%
        }
       
\begin{document}

\begin{abstract}
The heavy ion physics approach to global event characterization has led us to instrument the
forward region in the PHENIX experiment at RHIC. In heavy ion collisions this coverage yields a
measurement of the ``spectator" energy and its distribution about the beam direction. This energy flow
is the basis of event-by-event determination of the centrality and reaction plane which are key to 
analyzing particle production in heavy ion collisions. These same tools have also enabled a unique set 
of measurements on inelastic diffraction with proton, deuteron and gold ion beams in the PHENIX 
experiment. We present first new results on this topic and discuss briefly the opportunity for
diffractive physics with Heavy Ion beams at the LHC. 
\vspace{1pc}
\end{abstract}

\maketitle

\section{Introduction}

Like the LHC, the RHIC collider is a dual function machine providing collisions of
protons as well as heavy ions. Unlike the LHC, the experiments at RHIC have been designed completely with the goals of heavy ion physics in mind and then subsequently used also for measurements with (polarized) proton colliding beams. In many cases the choice of parameters in detector design have had interesting consequences for the types of measurement that can be carried out with pp collisions but nowhere is this so clear as in the case of forward instrumentation.

\subsection{Heavy Ion Goals}

	The primary goals of forward instrumentation at RHIC are
\begin{itemize}
\item{to measure absolute luminosity}
\item{to provide a truly minimum bias trigger}
\item{to characterize the initial conditions of individual events(reaction plane orientation, impact
parameter) outside the mid-rapidity region}
\end{itemize}
	
	The interaction region geometry, depicted in Fig.~1, provides a location at z=$\pm18$ m of maximum momentum dispersion where ``spectator fragments" from the colliding nuclei are resolved primarily according to their charge-to-mass ratio (Z/A). An intermediate impact parameter event is shown in Fig.~2. RHIC experiments were designed to characterize the reaction plane and impact parameter of individual events. The number of forward going ``spectator nucleons" leaving the collision is anti-correlated with the number of participants:
\begin{equation}
	N_{{\rm spectator}}+N_{{\rm participant}}=2\times A.
\end{equation}
	
\begin{figure}[htb]
\includegraphics[width=8.0cm]{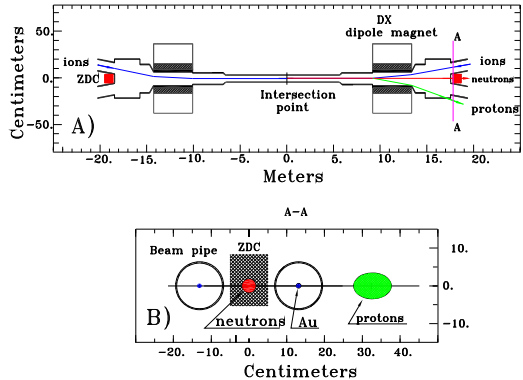}
\caption{RHIC interaction region layout.}
\label{zdc}
\end{figure}

In neutron rich ion species stored at RHIC (Au) and LHC (Pb) most of the information is carried by neutrons which are detected in the RHIC Zero Degree Calorimeters (ZDC's) \cite{Instrum} that were installed in each experiment. Protons are also measured in the PHENIX experiment using a large area hadron calorimeter recycled from AGS experiment E865. 

		The centrality measurement by spectators is partially diluted by a significant component of recombination/coalescence of the spectator nucleons (ie $n+p \rightarrow$ {\rm d}) which are neither detected in the ZDC or the FCAL. Even for central collisions $\sim1/2$ of the protons coalesce to higher mass fragments \cite{NA49}. So in PHENIX the participant number is sampled by the multiplicity in the 
$3\leq | \eta| \leq 4$ region and combined with ZDC energy to form the centrality algorithm.

	Early studies of RHIC collisions \cite{MCD} showed that spectator neutrons are detected in the ZDC with ${\rm eff_{ZDC}}\geq99.5\%$ in nuclear interactions at all centralities. Even in UltraPeripheral Collisions (UPC) where $R_{{\rm impact}}\geq 2 \times R_{{\rm Au}}$ , beam energy neutrons are evaporated (and detected in the ZDC's) due to the intense electromagnetic interactions between high-Z nuclei \cite{Baltz}.  The ZDC is therefore the ideal minimum biased event trigger for nuclear interaction physics. For UPC, the ZDC trigger (``tagged" UPC events-see below) is used to form ``rapidity gap" triggers and is the only practical way to access this physics.

	Because the dominant UPC process -mutual coulomb dissociation - is calculable with an
accuracy of $\sim 5\%$ \cite{Baltz} and cleanly measured \cite{MCD}, it is used as the basis for all absolute luminosity measurements in RHIC.
	
	The reaction plane orientation is best measured in the ZDC using a position sensitive scintillator array at a depth of $2\times\lambda_{{\rm Int}}$ in the calorimeter because of the large directed flow, $v_1$, among  nucleons at this rapidity \cite{Sorge}. Since the orientation is measured independently in both ZDC arms(left and right) the resolution can be directly derived and we find
\begin{eqnarray}
0.3&=&<cos(\psi_{meas}-\psi_{true})> \nonumber\\
&=&\sqrt{2}\cdot <cos(\psi_{right}-\psi_{left})> 
\end{eqnarray}
	
\begin{figure}[htb]
\includegraphics[width=7.5cm]{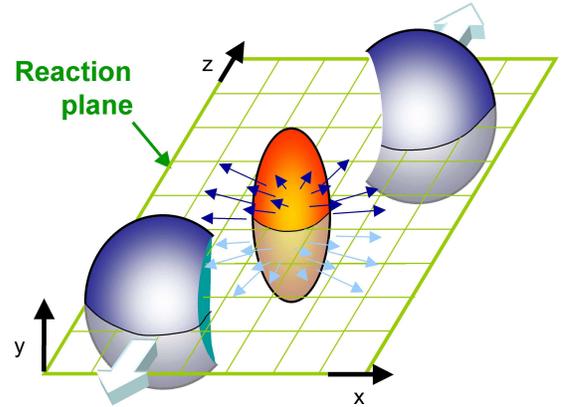}
\caption{Collision geometry in heavy ion collisions. Both magnitude and direction of the impact parameter are measured by forward ``spectator" remnants. }
\label{react}
\end{figure}

\section{{\it EM} Interactions of Heavy Ions}

	Ultraperipheral collisions become an important aspect of heavy ion physics at the large c.m. energies available at colliders. At the LHC, with Pb+Pb collisions at $\sqrt{S_{NN}}=5.5$ TeV, processes can be studied that are equivalent to 100 TeV quasi-real photon collisions with a stationary nuclear target \cite{white_erice} opening up a wide range of opportunities \cite{hot_topics}.
	
	The principal method for analyzing such collisions, the method of equivalent quanta, originated with Fermi \cite{fermi} who showed that interactions of electrons and $\alpha-$particles with atoms can be reduced to a convolution of an equivalent photon spectra and the known $\gamma-$A interaction cross sections. The method was extended to the extreme relativistic case by Weizsacker and Williams and has found many applications in high energy and nuclear physics. 
	
	At ion colliders the characteristic strength of interaction $\sim Z_1\times Z_2 \times \alpha$ is large so that multiple photon exchanges lead to the phenomenon of tagging. Additional photon exchanges can cause photo-nuclear breakup of one nucleus which emits a soft neutron \cite{Baur}  detected in the ZDC. Originally the tagging requirement was thought of as a penalty factor required to access this 
physics with a clean experimental trigger. However, as shown in Ref. \cite{Baur} the tagging fraction is correlated with impact parameter and strongly biases the tagged photon distribution toward higher equivalent energies. We now calculate tagged and untagged photon spectra separately, using this
nomenclature below.

\subsection{ Diffractive vector meson production at RHIC}

	During the $\sqrt{S_{NN}}=130$ GeV initial Au+Au run, STAR recorded data with a two-prong track requirement (consistent with diffractive $\rho\rightarrow\pi^+\pi^-$ decays) as well as a minibias ZDC trigger \cite{starrho}. The two data samples were examined for diffractive $\rho$ production as well as low mass $\gamma\gamma\rightarrow e^+e^-$candidates.
	
	The results are summarized in Table.~1, reproduced from Ref. \cite{starrho} where results are also compared with a calculation of the $\rho$ production cross sections for both the tagged and untagged cases \cite{BKN}. The cross sections (subscripts 0n and Xn denote 0 and 1-or more neutron tags, respectively) are in reasonable agreement with the calculation. Note that the experimental errors are smallest in the tagged sample.

\vspace*{0.3cm}
\begin{table}

\begin{tabular}{|l|c|c|}
\hline 
Cross Section    & STAR (mb) & ref. (mb) \\  
\hline
$\sigma^\rho_{{\rm xn,xn}}$ & $ 28.3 \!\pm\! 2.0  \!\pm\! 6.3$ &$  27$ \\
$\sigma^\rho_{{\rm 1n,1n}}$ &$ 2.8 \!\pm\! 0.5\!\pm\! 0.7$ & $ 2.6$ \\
\hline
$\sigma^{\rho{\rm (inc. overlap)}}_{xn,xn}$ & $ 39.7  \!\pm\! 2.8  \!\pm\! 9.7$ &$  - $ \\
$\sigma^\rho_{{\rm xn,0n}}$ & $95 \!\pm\! 60  \!\pm\! 25$ &$  - $ \\
$\sigma^\rho_{{\rm 0n,0n}}$   & $370\!\pm\!170 \pm 80$ & $-$ \\
\hline
$\sigma^\rho_{{\rm total}}$ & $460\!\pm\!220\!\pm\!110$ & $350$ \\
\hline
\end{tabular}
\caption{$\rho$ photoproduction results reproduced from Ref.~\cite{starrho}.}
\end{table}	

	During the most recent (2004) Au+Au run, PHENIX implemented a diffractive trigger sensitive to high mass $e^+e^-$ pairs using a ZDC tag and vetoing on events with activity in the $3\leq | \eta| \leq 4$ (=BBC) region.
The PHENIX experiment has excellent identification capabilities for high momentum electrons due to the highly segmented ($\Delta\eta\times\Delta \Phi=0.01\times0.01$) Electromagnetic (EM) Calorimeter and Ring Imaging Cerenkov counters. The diffractive $e^+e^-$ trigger also required at least one cluster of calorimeter energy with greater than 0.8 GeV. The full trigger is
\begin{equation}
{\rm UPC=ZDC( or ) \cdot BBC(\it{not}and)\cdot EM}.
\end{equation} 

	This loose rapidity gap trigger ({\it ie} a leading neutron and no activity in at least one BBC arm) is very effective in heavy ion collisions and had a rate of $\leq 0.5\%$ of the min bias rate. A total of 8.5 M UPC triggers were recorded. Electron candidates were reconstructed requiring consistent EM calorimeter energy and magnetic spectrometer momentum as well as a Cerenkov match. The resultant $e^+e^-$ invariant mass spectrum \cite{david} is shown in Fig.~3. All candidates have a net $p_t$ of $\leq 100$ MeV, consistent with diffractive photo-production and the expected Au nuclear form factor. This should be compared to the $\overline{p_t}\geq 1$ GeV/c we find in non-diffractive $p+p\rightarrow J/\psi$+X data . 
	
	It should be noted that this spectrum is highly preliminary since, at the time of this initial pass through the data, there has been very little prior experience in PHENIX reconstructing events with such low multiplicities ({\it ie} where the interaction vertex is not well constrained). Our calculation of the expected J/$\psi$ and $e^+e^-$ continuum yield is higher than observed by about a factor of 2. Nevertheless, even at this very early stage in the analysis, it is clear that a diffractive signal of high mass pairs has been seen and the capability for small cross section diffractive physics has been demonstrated.

\begin{figure}[htb]
\includegraphics[width=7.5cm]{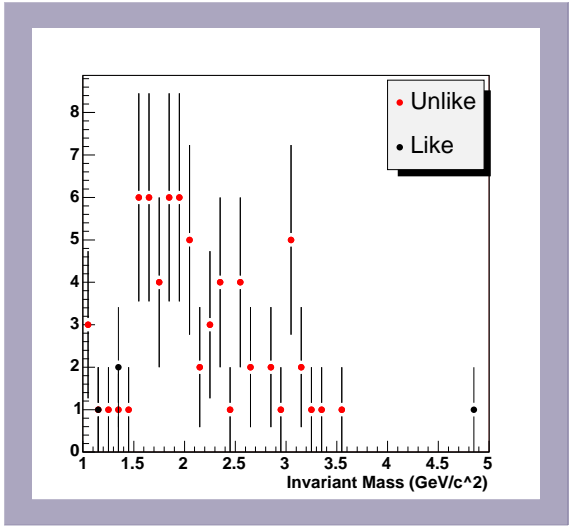}
\caption{Preliminary PHENIX diffractive $e^+e^-$ mass spectrum.}
\label{eemass}
\end{figure}

\section{Au+d$\rightarrow$ Au$+n+p$}

	Immediately following the first high luminosity Au+Au run RHIC operated as a d+Au collider (at $\sqrt{S_{NN}}=200$ GeV). The primary purpose of this run was to directly measure the contribution of initial state nuclear effects on inclusive particle spectra observed in d-Au$\rightarrow \pi^0$+X for example. In order to measure inclusive particle cross sections, it is critical to calibrate the d-Au luminosity. Also, since particle spectra are compared for an equivalent number of binary collisions, derived from a Glauber model calculation, it is useful to validate this calculation by confirming the corresponding minimum bias cross section.
	
		During the d+Au run STAR and PHENIX ran with different min.bias triggers:
\begin{eqnarray}
{\rm min.bias(STAR)}&=&{\rm ZDC_{Au-beam}}(\geq 15 {\rm GeV})\nonumber\\
{\rm min.bias(PHENIX)}&=&{\rm BBC_{left}\cdot BBC_{right}}
\end{eqnarray}
which, as pointed out in Ref. \cite{boris}, have different acceptances for $NN$ collisions with a rapidity gap and hence different cross sections. 

	In PHENIX we have measured directly the inelastic d+Au cross section by normalizing to the theoretical calculation \cite{vogt} for deuteron photodissociation of $\sigma_{{\rm diss}}=1.38 \pm 0.07$ b.
	PHENIX recorded 12 M events with three different triggers enabled: the 2 inelastic triggers of eqn.~4 in ``or" with a trigger requiring ${\rm ZDC_{d-beam}}\geq15$ {\rm GeV}. This last trigger is $100\%$ efficient for deuteron breakup events since, with a reasonable deuteron input wavefunction, all neutrons fall within the ZDC aperture. Dissociation events have a particularly clean signature in the PHENIX forward detectors since, in the d beam direction, there is an easily resolved 100 GeV energy peak simultaneously in both the ZDC and the proton calorimeter.  Also there is no activity in the central detector or the Au beam ZDC. 
	
	Ref.~\cite{boris} predicts a factor of 0.83 suppression of the min.bias(PHENIX) cross section relative to the naive Glauber value of $\sigma_{{\rm Glauber}}=2.33$ b. Instead we measure, applying a small correction for BBC efficiency:
	\begin{equation}
	\sigma_{\rm {PHENIX}(min.bias)}=2.39\pm 0.24 {\rm b},
	\end{equation}
	
	which differs by $\sim 2$ st.dev. from Ref. \cite{boris}. The error on this preliminary result will probably reach $\leq5 \%$ after further study.

\begin{figure}[htb]
\vspace{70pt}
\includegraphics[width=7.cm]{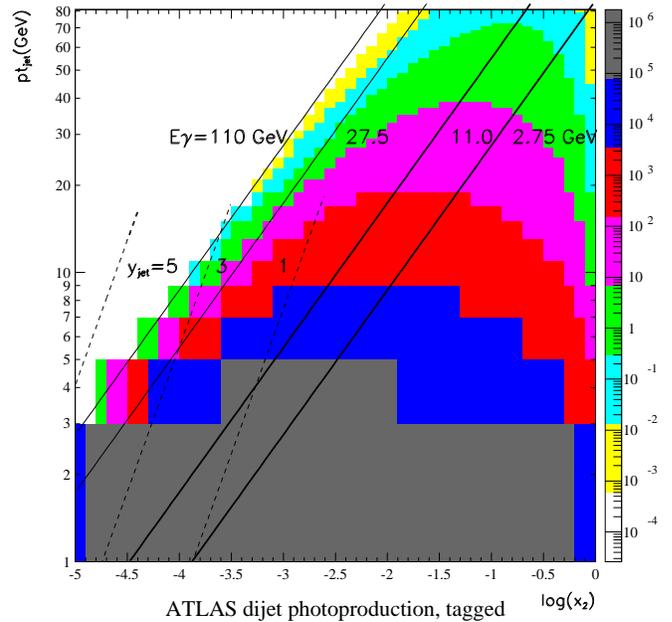}
\caption{Rates of photo-produced jet pairs from a 1 month run in ATLAS.}
\label{jetjet}
\end{figure}

\section{$pp\rightarrow n+X$}

	Large $x_{F}$ inclusive cross sections are relevant for analysis of super-GZK cosmic ray candidates and also for LHC luminometry, where this process will be used for beam diagnostics. At RHIC it has a special role in polarimetry since there is a large ($\sim 10\%$) left-right asymmetry of the neutron angular distribution relative to the up-down polarization of the incident proton. Also a large ($\sim 1 \%$)  $pp\rightarrow nn$+X signal is observed which is not reproduced by most event generators ({\it ie} DPMJETIII). This signal is the basis of all RHIC $pp$ luminosity monitoring. Further study of these forward physics processes is now under way.
	
\section{Photoproduction with Heavy Ions at the LHC}

	The ATLAS experiment is now constructing ZDC's to be installed in the TAN absorber at $\pm 140$ m from the IP1 interaction region. Full geant simulation of this detector with $\sqrt{S_{NN}}=5.5$ TeV Pb+Pb collisions has demonstrated a similar energy resolution to that observed at RHIC.
	
	The ATLAS ZDC's will be the primary trigger for diffractive physics and will be combined with low-$p_t$ jet or secondary vertex triggers. A likely area of research is the study of photo-produced dijets, $\gamma-$jet pairs and heavy quarks ($Q\overline{Q}$) and also the corresponding diffractive photo-production processes (which have a rapidity gap in {\it both} beam directions). We have carried out \cite{ramona} a calculation of rates for the above processes and the jet-jet case is presented in Fig.~4 for a 1 month run at nominal LHC Pb-Pb luminosities. Even with a jet threshold of $\geq 10-15$ GeV the total event rate exceeds $\sim10^5$. We are now studying the ATLAS capability for accessing low $p_t$ jets in events with the very low underlying multiplicities expected for these diffractive processes. It is likely that soft heavy quark production can be tagged through a secondary vertex at lower $p_t$. This would probe the lowest $x_2$ parton distributions(abscissa in Fig.~4)- possibly down to $x_2\sim 10^{-4}$. We also indicate in Fig.~4 the equivalent incident photon energy and the average rapidity($y_{{\rm jet}}$) corresponding to each bin. For the most part, the produced jets are very central and it should be possible to require a large rapidity gap along the direction of the beam emitting the photon. The ATLAS forward calorimeters are unique in providing sensitivity up to $|\eta|<5.0$ for $E_T\geq 2.0$ GeV and this makes it the most promising experiment to carry out  rapidity gap physics with heavy ions.
	
\section{Acknowledgements}

	I am grateful to the organizers of ``Diffraction 2004" for a very stimulating and pleasant meeting.

\end{document}